\documentclass[aps,prd,superscriptaddress,twoside,twocolumn,nofootinbib,%
10pt,showpacs,floatfix]{revtex4-2}

\usepackage{amsmath,amssymb}
\usepackage{graphicx,bm}
\usepackage{ulem}
\usepackage{slashed}
\usepackage{dcolumn}
\usepackage{epsfig}
\usepackage{multirow}

\allowdisplaybreaks

\begin{document}


\title{Tetraquark mixing model is superior to meson molecules in explaining two light-meson nonets}


\author{Hungchong Kim}%
\email{bkhc5264@korea.ac.kr}
\affiliation{Center for Extreme Nuclear Matters, Korea University, Seoul 02841, Korea}

\author{K. S. Kim}%
\email{kyungsik@kau.ac.kr}
\affiliation{School of Liberal Arts and Science, Korea Aerospace University, Goyang, 412-791, Korea}

\date{\today}


\begin{abstract}
In this work, we compare the tetraquark mixing model and meson molecules in describing the two physical nonets in the $J^P=0^+$ channel,
the light nonet [$a_0 (980)$, $K_0^* (700)$, $f_0 (500)$, $f_0 (980)$]
and the heavy nonet [$a_0 (1450)$, $K_0^* (1430)$, $f_0 (1370)$, $f_0 (1500)$].
In particular, we focus on whether successful aspects of the tetraquark mixing model that apply to all members of each nonet
can be reproduced from a model of meson molecules.
By combining two mesons in the lowest-lying pseudoscalar nonet, we construct SU$_f$(3) molecular nonets that can be tested for the two physical nonets.
This molecular approach can make two flavor nonets just as the tetraquark mixing model
but this model has some difficulties in describing the universal features of the two nonets
such as mass splitting between the two nonets, and enhancement or suppression of the coupling strengths
of the two nonets into two pseudoscalar mesons.
We also compare the fall-apart modes of the tetraquark mixing model and the two-meson modes from the molecular model.
A clear distinction can be seen by the two-pion modes in the isovector resonances.
The two-pion modes appear in the molecular model, but not in the tetraquark mixing model.
The absence of the two-pion modes is supported by the experimental decay modes of the isovector resonances.
\end{abstract}


\maketitle

\section{Introduction}

\vspace{20pt}
Multiquarks are an important topic to be studied in hadron physics.
Currently, there are many candidates for multiquarks in hadron spectroscopy.
There are several candidates for hidden-charm tetraquarks, including $\chi_{c1} (3872)$, $X^{\pm} (4020)$, $\chi_{c1} (4140)$,
$Z_c (3900)$~\cite{Belle03, BESIII:2013ouc, LHCb:2016axx, Xiao:2013iha}, and the
doubly-charmed tetraquark candidate $T_{cc}^+$~\cite{LHCb:2021auc,LHCb:2021vvq}. Additionally, candidates for hidden-charm pentaquarks, such as
$P_c (4312)$, $P_c (4440)$ and $P_c(4457)$ have been reported in Ref.~\cite{LHCb:2015yax,LHCb:2019kea}.
In the light quark sector ($q=u,d,s$), there have been long-standing candidates for tetraquarks, the light nonet
consisting of $a_0 (980)$, $K_0^* (700)$, $f_0 (500)$, and $f_0 (980)$~\cite{Jaffe77a, Jaffe77b, Jaffe04}.
The heavy nonet composed of $a_0 (1450)$, $K_0^* (1430)$, $f_0 (1370)$, and $f_0 (1500)$ are also
expected to be tetraquarks generated by the tetraquark mixing model~\cite{Kim:2016dfq, Kim:2017yvd, Kim:2018zob,Kim:2018zob,Lee:2019bwi,Kim:2017yur,Kim:2022qfj,Kim:2019mbc}.

Perhaps a major difficulty in confirming these candidates as multiquarks is that they can also be described as
composite systems of hadrons, which are often referred to as hadronic molecules~\cite{Guo:2017jvc}.
In this description, they are treated as two color-singlets,
such as meson-meson bound systems, meson-baryon systems, or states that are dynamically generated from two hadrons.
Specifically, the $\chi_{c1}$(3872) observed by the Belle Collaboration~\cite{Belle03} could be a tetraquark with the flavor structure
of $cq\bar{c}\bar{q}~ (q=u,d)$~\cite{Maiani:2004vq, Kim:2016tys} or it could be a meson molecular state composed
of $D\bar{D}^*$~\cite{Tornqvist:2004qy, Tornqvist:1993ng}.
The $P_c (4312)$, $P_c (4440)$, $P_c$(4457) resonances observed in Ref.~\cite{LHCb:2015yax,LHCb:2019kea}
may be the pentaquarks with a structure of $uudc\bar{c}$. Or they could be hadronic molecules,
$\Sigma_c \bar{D}$ ($J^P=1/2^-$), $\Sigma_c \bar{D}^*$ ($J^P=3/2^-$), $\Sigma_c \bar{D}^*$ ($J^P=1/2^-$), respectively~\cite{Du:2019pij,Xiao:2019mvs}.
The $d^*(2380)$ resonance reported in Ref.~\cite{WASA-at-COSY:2011bjg} may be a hexaquark state~\cite{Kim:2020rwn} or it could be
a $\Delta\Delta$ molecular state as predicted by Dyson and Xuong~\cite{Dyson:1964xwa}.
A similar confusion exists in the light quark system.
The light and heavy nonets discussed above may be tetraquarks realized by a mixture of the two tetraquark types~\cite{Kim:2016dfq, Kim:2017yvd, Kim:2018zob,Kim:2018zob,Lee:2019bwi,Kim:2017yur,Kim:2022qfj,Kim:2019mbc}.
But at the same time, some members of the light nonet, such as $a_0 (980)$ and $f_0 (980)$, can be interpreted as
molecular states like $K\bar{K}$ or dynamically generated from $\pi\eta$ or $K\bar{K}$~\cite{Weinstein:1990gu, Branz:2007xp, Branz:2008ha, Janssen:1994wn}.
The isoscalar resonance, $f_0(500)$, may be a meson molecule composed of $\pi\pi$~\cite{Ahmed:2020kmp}.
In the heavy nonet, the $f_0(1370)$ can be a $\rho\rho$ molecule~\cite{Molina:2008jw}.

So it appears that most candidates for multiquarks can be described also by hadronic molecules.
But multiquarks and hadronic molecules are different states clearly distinguished by their color configurations.
In color space, hadronic molecules are composed of two color-singlets while multiquarks, as they are constructed from
colorful constituents like diquarks,
have hidden color configurations in addition to the molecular configuration.
With this difference in mind, we need to choose appropriate candidates for multiquarks and contemplate
how they can be confirmed as multiquarks clearly distinguished from hadronic molecules.

Promising candidates in this regard could be the two nonets in Particle Data Group(PDG)~\cite{PDG22}:
the light nonet [$a_0 (980)$, $K_0^* (700)$, $f_0 (500)$, $f_0 (980)$] and
the heavy nonet [$a_0 (1450)$, $K_0^* (1430)$, $f_0 (1370)$, $f_0 (1500)$].
All members of each nonet, which seem to form an SU(3)$_f$ flavor nonet, $\bm{9}_f$,
are relatively well known experimentally with many physical properties that can be utilized to investigate their nature as multiquarks.
For identifying multiquarks, the two nonets are certainly advantageous over the other candidates in the charm quark sector
for which family members of the SU(3) multiplets have yet been discovered.
According to the tetraquark mixing model~\cite{Kim:2016dfq, Kim:2017yvd, Kim:2018zob,Kim:2018zob,Lee:2019bwi,Kim:2017yur,Kim:2022qfj,Kim:2019mbc},
the two nonets are tetraquarks created by mixing two types of tetraquarks that separately form a flavor nonet.
So the two nonets must be treated together.
This tetraquark mixing model is phenomenologically successful in various aspects to be discussed in Sec.~\ref{tmm}.
We emphasize that the successful aspects are universal to all members of each nonet and not limited to a few members.

To establish this tetraquark mixing model, it is important to test alternative models such as meson molecules for the two nonets.
As we have already mentioned, some members of the two nonets can be described as meson
molecules~\cite{Weinstein:1990gu, Branz:2007xp, Branz:2008ha, Janssen:1994wn,Molina:2008jw}.
If the two nonets separately form a flavor nonet,
the remaining members of the nonets are also expected to be meson molecules as they
can be generated via SU(3)$_f$ rotations.  This aspect can be investigated by combining
two pseudoscalar (PS) mesons of ($\pi,K,\eta,\eta^\prime$).
Since the lowest-lying pseudoscalar mesons form a flavor nonet, the two-meson states constructed from them
can make various multiplets.
We then ask whether this
molecular model can create two flavor nonets that can describe the two physical nonets appropriately.
Does this model also reproduce the successful aspects of the tetraquark mixing model.
From these considerations, we can identify the differences between the meson molecular model and the tetraquark mixing model,
and decide which model is more realistic for the two nonets in PDG.
This type of research eventually helps to determine whether the two nonets in PDG are tetraquarks or not.

Another perspective that differentiates the two models can be seen by examining the two-meson modes from both models.
Tetraquarks in the mixing model take the diquark-antidiquark form, $(qq)(\bar{q}\bar{q})$,
which can be rearranged into two pairs of quark-antiquark, $(q\bar{q})(q\bar{q})$.
From this rearrangement, one can see that the tetraquarks have the two-meson components consisting of two color-singlets, $(q\bar{q})_{\bm{1_c}} (q\bar{q})_{\bm{1_c}}$,
and the hidden color components like $[(q\bar{q})_{\bm{8_c}} (q\bar{q})_{\bm{8_c}}]_{\bm{1_c}}$.
It is quite likely that the two-meson components can inadvertently lead us to identify the tetraquarks as meson molecules.
On the other hand, in meson molecules, the two nonets are built solely from two-meson states
that are combined according to the SU$_f$(3) symmetry.
The two-meson components in the tetraquark mixing model, when viewed in flavor space,
are in principle different from the two-meson modes of meson molecules.
Tetraquarks in the mixing model form a flavor nonet, $\bm{9}_f~(=\bm{1}_f\oplus\bm{8}_f)$.
The two-meson components in this case, therefore, are restricted to specific combinations that are
governed by the original flavor structure, $\bm{9}_f$, of the tetraquarks.
In contrast, the two-meson modes in the molecular model, when constructed by two PS mesons,
can have various meson combinations allowed by $\bm{9}_f\otimes \bm{9}_f\Rightarrow \bm{9}$
where $\bm{9}_f$ denotes a nonet of the lowest-lying PS meson.
The resulting nonet does not necessarily have the same meson combinations as those from
the tetraquark wave functions. Eventually, the experimental decay modes of the two nonets
can be used to determine which model is more realistic for the two nonets.

This paper is organized as follows. In Sec.~\ref{tmm}, we review the tetraquark mixing model
explaining its structure and successful aspects.  Two-meson modes from the tetraquark wave functions will be presented also.
In Sec.~\ref{mm}, we construct two SU$_f$(3) nonets from two PS mesons using a tensor method.
In Sec.~\ref{comparison}, two-meson modes from the two approaches will be compared.
We then discuss the phenomenological limitations of the meson molecular model in describing the two nonets in PDG.

\section{tetraquark mixing model and two-meson modes }
\label{tmm}

In this section, we review the tetraquark mixing model that has been constructed for the two nonets in PDG~\cite{Kim:2016dfq, Kim:2017yvd, Kim:2018zob,Kim:2018zob,Lee:2019bwi,Kim:2017yur,Kim:2022qfj,Kim:2019mbc}.
In the tetraquark mixing model, the two nonets are treated as tetraquarks produced by the mixture of two tetraquark types,
denoted as $|000\rangle$, $|011 \rangle$.  The $|000\rangle$ type represents the spin-0 tetraquarks constructed
by combining the {\it spin-0} diquark of the color and flavor structures ($\bar{\bm{3}}_c, \bar{\bm{3}}_f$)
and its antidiquark. The $|011\rangle$ type also represents the spin-0 tetraquarks but
constructed by the {\it spin-1} diquark of the structure ($\bm{6}_c, \bar{\bm{3}}_f$) and its antidiquark.
The two tetraquark types differ by color and spin configurations and, because of this, they strongly mix through the color-spin interaction,
$V_{CS} \sim \sum_{i < j}  \frac{\lambda_i \cdot \lambda_j J_i\cdot J_j}{m_i^{} m_j^{}}$~\cite{Kim:2016dfq, Kim:2017yvd}.
This strong mixing in effect causes a huge mass gap between the two nonets.
The physical two nonets can be identified by the linear combinations,
\begin{eqnarray}
|\text{Heavy~nonet} \rangle &=& -\alpha | 000 \rangle + \beta |011 \rangle \label{heavy}\ ,\\
|\text{Light~nonet} \rangle~ &=&~~\beta | 000 \rangle + \alpha |011 \rangle \label{light}\ ,
\end{eqnarray}
that diagoanlize the color-spin interaction, $V_{CS}$.
The mixing parameters are $\alpha\approx \sqrt{{2}/{3}}$, $\beta\approx 1/\sqrt{3}$ fixed also by the diagonalization~\cite{Kim:2017yvd}.

The two tetraquark types, $|000\rangle$ and $|011\rangle$, have the same flavor configuration. The two types separately form
a flavor nonet, $\bm{9}_f$, as both are constructed by combining the diquark with $\bar{\bm{3}}_f$ and its antidiquark with $\bm{3}_f$.
Because of this, both types generate the ``inverted mass ordering''~\footnote{The inverted mass ordering refers to the ordering,
$M[a_0(980)]>M[K^*_0(700)]>M[f_0(500)]$, which is
inverted from the mass ordering expected from a two-quark picture, $M[a_0(980)]<M[K^*_0(700)]<M[f_0(500)]$.
Therefore, according to Refs.~\cite{Jaffe77a,Jaffe77b,Jaffe04}, this inverted mass ordering is crucial evidence indicating that the light
nonet members are tetraquarks.}
that are clearly satisfied by the members of the light nonet
and marginally by the members of the heavy nonet~\cite{Kim:2016dfq, Kim:2017yvd,Kim:2018zob}.

The tetraquark mixing model, which is represented by Eqs.~(\ref{heavy}),(\ref{light}),
has several successful aspects in describing the two nonets in PDG.
First, the tetraquark mixing model explains relatively well the mass gap, $\Delta M$, between the two nonets
by the hyperfine mass splitting, $\Delta \langle V_{CS} \rangle$~\cite{Kim:2016dfq, Kim:2017yvd}.
Second, the mixing model makes huge hyperfine mass for the light nonet,
approximately $\langle V_{CS} \rangle\approx -500$ MeV, which can substantially reduce the mass of the light nonet.
This explains qualitatively why the members of the light nonet, despite being tetraquarks, can have masses below
1 GeV~\cite{Kim:2016dfq,Kim:2017yvd,Kim:2018zob,Kim:2019mbc}.
At the same time, the mixing model produces small hyperfine mass for the heavy nonet, approximately $\langle V_{CS} \rangle\approx -20$ MeV.
This can explain why the members of the heavy nonet have masses not far from $4m_q$, four times that of the constituent quark mass.

The most striking prediction of the mixing model is that the coupling strengths of the two nonets into two PS mesons
are enhanced in the light nonet but suppressed in the heavy nonet~\cite{Kim:2017yur,Kim:2022qfj}.
This prediction comes from the fact that $|000\rangle$ or $|011\rangle$ can have two-meson components when their wave functions,
originally written in diquark-antidiquark form, are rearranged into two pairs of quark-antiquark.
Tetraquarks can fall-apart into two PS mesons through the two-meson components.
The associate coupling strengths can be calculated by collecting the recombination factors from color, spin, and flavor space in the rearrangement.
The flavor recombination factors should be the same for both, $|000\rangle$ and $|011\rangle$, as the two tetraquark types
have the same flavor configuration. But the color and spin recombination factors are different
because $|000\rangle$ and $|011\rangle$ have different color and spin configurations.
What is interesting is that $|000\rangle$ and $|011\rangle$ have opposite signs in the heavy nonet, Eq.~(\ref{heavy}),
while they have the same sign in the light nonet, Eq.~(\ref{light}).
Due to the difference in relative signs, the two-meson modes partially cancel out in the heavy nonet, but add up in the light nonet.
This is precisely the reason why the couplings into two PS mesons are enhanced in the light nonet but suppressed
in the heavy nonet~\cite{Kim:2017yur,Kim:2022qfj}.

To show this, we explicitly calculate the two-meson modes in the tetraquark mixing model by
rearranging $|000\rangle$ and $|011\rangle$ into two pairs of quark-antiquark.
The two-meson modes are as follows.
\begin{widetext}
\begin{eqnarray}
&&\text{\underline{Two-meson modes of the light nonet}:} \nonumber \\
&&K_0^{*+}(700): \left (\frac{\beta}{\sqrt{12}}+\frac{\alpha}{\sqrt{2}}\right )\frac{1}{2}\Big\{ \pi^+K^0 +K^0\pi^+ + \frac{1}{\sqrt{2}}(K^+\pi^0 + \pi^0 K^+)
- \frac{1}{\sqrt{6}}(K^+ \eta_8 + \eta_8 K^+) \nonumber \\
&&~~~~~~~~~ - \frac{1}{\sqrt{3}}(K^+\eta_1 +\eta_1 K^+)\Big\} \label{doublet700}\ ,\\
&&a_0^+(980):  \left (\frac{\beta}{\sqrt{12}}+\frac{\alpha}{\sqrt{2}}\right ) \frac{1}{2}\Big\{ \bar{K}^0 K^+ + K^+ \bar{K}^0
+ \sqrt{\frac{2}{3}}\left (\eta_8 \pi^+ + \pi^+ \eta_8\right )
- \frac{1}{\sqrt{3}}\left (\eta_1 \pi^+ + \pi^+ \eta_1 \right ) \Big\}\label{isovector980}\ ,\\
&&f_0(500):\left (\frac{\beta}{\sqrt{12}}+\frac{\alpha}{\sqrt{2}}\right )
\Bigg\{\frac{1}{3}\left[(a+\sqrt{2}b)\eta_1\eta_1+(\frac{a}{\sqrt{2}}-\frac{b}{2})\eta_1\eta_8+(\frac{a}{\sqrt{2}}-\frac{b}{2})\eta_8\eta_1
+(\frac{a}{2}-\sqrt{2}b)\eta_8\eta_8\right]\nonumber \\
&&~~~~~~~~~-\frac{a}{2}\bm{\vec{\pi}}\cdot\bm{\vec{\pi}}-\frac{b}{2\sqrt{2}}\left[\overline{K} K +(\overline{K} K)^\dagger\right]\Bigg\}\label{500mode}\ ,\\
&&f_0(980) :\left (\frac{\beta}{\sqrt{12}}+\frac{\alpha}{\sqrt{2}}\right )
\Bigg\{\frac{1}{3}\left[(\sqrt{2}a -b)\eta_1\eta_1 - (\frac{a}{2}+\frac{b}{\sqrt{2}})\eta_1\eta_8 - (\frac{a}{2}+\frac{b}{\sqrt{2}})\eta_8\eta_1
-(\sqrt{2}a+\frac{b}{2})\eta_8\eta_8\right]\nonumber \\
&&~~~~~~~~~+\frac{b}{2}\bm{\vec{\pi}}\cdot\bm{\vec{\pi}}-\frac{a}{2\sqrt{2}}\left[\overline{K} K +(\overline{K} K)^\dagger\right]\Bigg\}\label{980mode}\ .\\
\nonumber \\
&&\text{\underline{Two-meson modes of the heavy nonet}:} \nonumber \\
&&K_0^{*+}(1430):  \left(-\frac{\alpha}{\sqrt{12}}+\frac{\beta}{\sqrt{2}}\right) \frac{1}{2}\Big\{ \pi^+K^0 +K^0\pi^+ + \frac{1}{\sqrt{2}}(K^+\pi^0 + \pi^0 K^+)
- \frac{1}{\sqrt{6}}(K^+ \eta_8 + \eta_8 K^+) \nonumber \\
&&~~~~~~~~~ - \frac{1}{\sqrt{3}}(K^+\eta_1 +\eta_1 K^+)\Big\}\label{doublet1430}\ ,\\
&&a_0^+(1450):  \left(-\frac{\alpha}{\sqrt{12}}+\frac{\beta}{\sqrt{2}}\right) \frac{1}{2}\Big\{ \bar{K}^0 K^+ + K^+ \bar{K}^0
+ \sqrt{\frac{2}{3}}\left (\eta_8 \pi^+ + \pi^+ \eta_8\right )
- \frac{1}{\sqrt{3}}\left (\eta_1 \pi^+ + \pi^+ \eta_1 \right ) \Big\}\label{isovector1450} \ ,\\
&&f_0(1370):\left (-\frac{\alpha}{\sqrt{12}}+\frac{\beta}{\sqrt{2}}\right )
\Bigg\{\frac{1}{3}\left[(a+\sqrt{2}b)\eta_1\eta_1+(\frac{a}{\sqrt{2}}-\frac{b}{2})\eta_1\eta_8+(\frac{a}{\sqrt{2}}-\frac{b}{2})\eta_8\eta_1
+(\frac{a}{2}-\sqrt{2}b)\eta_8\eta_8\right]\nonumber \\
&&~~~~~~~~~-\frac{a}{2}\bm{\vec{\pi}}\cdot\bm{\vec{\pi}}-\frac{b}{2\sqrt{2}}\left[\overline{K} K +(\overline{K} K)^\dagger\right]\Bigg\}\label{1370mode}\ ,\\
&&f_0(1500): \left (-\frac{\alpha}{\sqrt{12}}+\frac{\beta}{\sqrt{2}}\right )
\Bigg\{\frac{1}{3}\left[(\sqrt{2}a -b)\eta_1\eta_1 - (\frac{a}{2}+\frac{b}{\sqrt{2}})\eta_1\eta_8 - (\frac{a}{2}+\frac{b}{\sqrt{2}})\eta_8\eta_1
-(\sqrt{2}a+\frac{b}{2})\eta_8\eta_8\right]\nonumber \\
&&~~~~~~~~~+\frac{b}{2}\bm{\vec{\pi}}\cdot\bm{\vec{\pi}}-\frac{a}{2\sqrt{2}}\left[\overline{K} K +(\overline{K} K)^\dagger\right]
\Bigg\}\label{1500mode}\ .
\end{eqnarray}
\end{widetext}
Here, we have introduced shorthand notations to denote that
\begin{eqnarray}
&&\bm{\vec{\pi}}\cdot\bm{\vec{\pi}}=\pi^+\pi^-+\pi^-\pi^+ + \pi^0\pi^0\ ,\\
&&\overline{K}K=K^-K^+ + \bar{K}^0K^0 \ ,\\
&&(\overline{K}K)^\dagger=K^+K^- + K^0\bar{K}^0\ .
\end{eqnarray}
The flavor mixing parameters, $a,b$, can be fixed according to three different scenarios depending on how we treat the
flavor mixing as in Ref.~\cite{Kim:2017yvd}.
Two-vector modes have not been specified here because most nonet members are too light to decay into two vector mesons. Two-vector modes are not measurable
mostly so they are not useful for our comparison study with the molecular model.

Two-meson modes specified for each resonance in Eqs.~(\ref{doublet700})~$\cdot\cdot\cdot$~(\ref{1500mode})
are possible fall-apart modes into PS mesons predicted from the tetraquark mixing model.
Most of them can be seen as experimental decay modes of the two nonets in PDG if the decays are kinematically allowed.
Also we want to stress that the two-meson modes,
i.e. two PS or two vectors, do not represent the entire wave function of the
corresponding resonance. There are additional hidden color components [16,17] that
can genuinely distinguish tetraquarks from hadronic molecules.

Note, the coefficient of each two-meson mode in Eqs.~(\ref{doublet700})~$\cdot\cdot\cdot$~(\ref{1500mode}) can be identified as
the coupling strength of the corresponding resonance into those two mesons.
For example, the coefficient of $\bar{K}^0 K^+$ in Eq.~(\ref{isovector1450}) can be obtained
by $\langle \bar{K}^0 K^+ |a_0^+(1450)\rangle$ that defines the coupling strength between the $a_0^+(1450)$ and $\bar{K}^0 K^+$.
The coefficients of the light nonet in Eqs.~(\ref{doublet700})~$\cdot\cdot\cdot$~(\ref{980mode})
have the common overall factor, $\frac{\beta}{\sqrt{12}}+\frac{\alpha}{\sqrt{2}}\approx 0.744$,
obtained from color and spin recombining factors, while the heavy nonet
in Eqs.~(\ref{doublet1430})~$\cdot\cdot\cdot$~(\ref{1500mode}) has the overall factor, $ -\frac{\alpha}{\sqrt{12}}+\frac{\beta}{\sqrt{2}}\approx 0.173$.
This clearly shows that the coupling strengths are universally enhanced
in the light nonet but suppressed in the heavy nonet, due to the relative sign differences
originating from Eqs.~(\ref{heavy}),(\ref{light}).
After taking out the overall factors, the rest coefficient in each resonance is normalized to the unity.
We also notice that both nonets have the same two-meson modes
as expected from the fact that $|000\rangle$ and $|011\rangle$ have the same flavor configuration.
In this sense, the enhancement or suppression of the couplings in the tetraquark mixing model is a general consequence
that universally applies to all members of the two nonets.

As reported in Ref.~\cite{Kim:2022qfj}, this prediction can be verified qualitatively
by experimental partial decay widths extracted from PDG~\cite{PDG22}.
To explain this briefly, let us write partial decay width for a decay process as
\begin{eqnarray}
\Gamma_{partial} = G^2 \Gamma_{kin}\label{formula}\ ,
\end{eqnarray}
where $G$ is the coupling strength and $\Gamma_{kin}$ is so called ``kinematical partial width'',
which depends only on kinematical factors in the decay process.
Kinematically, the heavy nonet, as its mass is much heavier,
is expected to have much larger partial width than the light nonet. This mean, for $\Gamma_{kin}$, we should have
\begin{eqnarray}
\Gamma_{kin}(\text{light nonet}) \ll \Gamma_{kin}(\text{heavy nonet})\label{kin width}\ .
\end{eqnarray}
However, the experimental partial width, $\Gamma_{exp}$, extracted from PDG~\cite{PDG22}, shows an opposite tendency~\cite{Kim:2022qfj},
\begin{eqnarray}
\Gamma_{exp}(\text{light nonet}) \geq \Gamma_{exp}(\text{heavy nonet})\label{trend}\ .
\end{eqnarray}
Since the partial width can be calculated by Eq.~(\ref{formula}),
this opposite tendency in the experimental partial width, Eq.~(\ref{trend}),
must be reproduced if we multiply the corresponding $G^2$ on both sides of Eq.~(\ref{kin width}).
Only way to reproduce Eq.~(\ref{trend}) is to have the coupling strength of
the light nonet much larger than the coupling of the heavy nonet,
\begin{eqnarray}
G^2(\text{light nonet}) \gg G^2(\text{heavy nonet})\label{coupling strength}\ .
\end{eqnarray}
In addition, this result can provide an another explanation why $f_0(500)$, $K^*_0(700)$ have such large decay widths.

All the successful aspects of the tetraquark mixing model come rather naturally from the
fact that the two nonets are created by mixing the two tetraquark types.
None of the successful aspects can be obtained without mixing.
In particular, the enhancement or the suppression of the coupling strengths critically depends on
the fact that the two tetraquark types add to or partially cancel out in
Eqs.~(\ref{heavy}),(\ref{light}) that have been created from the mixing.
Moreover, this interesting fact is supported by the experimental partial widths
clearly indicating that the two nonets should not be treated separately.
This makes the tetraquark mixing model special and hopefully differentiates it from other models in describing the two nonets.

\section{Meson molecular states}
\label{mm}

Next, we examine whether the two nonets in PDG can be described also by models other than tetraquarks.
Specifically, we try to construct two-meson modes
like those in Eqs.~(\ref{doublet700})$\cdot\cdot\cdot$~(\ref{1500mode}) from a meson molecular model.
Since all the modes that we have considered in Eqs.~(\ref{doublet700})$\cdot\cdot\cdot$~(\ref{1500mode}) are fall-apart modes into
two PS mesons, we construct molecular states from the PS mesons
and see how the resulting combinations are different from those in the tetraquark mixing model.

The lowest-lying PS nonet is composed of a singlet, $\bm{1}_f=\eta_1$, and an octet, $\bm{8}_f$, that can be written
in a matrix form as,
\begin{eqnarray}
P^i_j &=&
\begin{bmatrix}
P^1_1 & P^2_1 & P^3_1 \\
P^1_2 & P^2_2 & P^3_2 \\
P^1_3 & P^2_3 & P^3_3
\end{bmatrix}\nonumber \\
\nonumber \\
&=&
\begin{bmatrix}
\frac{1}{\sqrt{2}}\pi^0+\frac{1}{\sqrt{6}}\eta_8 & \pi^+ & K^+ \\
\pi^- &-\frac{1}{\sqrt{2}}\pi^0+\frac{1}{\sqrt{6}}\eta_8 & K^0 \\
K^- &\bar{K}^0 &-\sqrt{\frac{2}{3}}~\!\eta_8
\end{bmatrix}\label{pij}\ .
\end{eqnarray}
This octet, $P^i_j$, is traceless ($P^i_i=0$) and normalized according to
\begin{eqnarray}
[P^i_j, P^k_l]=\delta^{ik}\delta_{jl}-\frac{1}{3}\delta^i_j\delta^k_l\label{normalization1}\ .
\end{eqnarray}
In the SU$_f$(3) symmetric limit, $\eta_1=\eta^\prime$ and $\eta_8=\eta$.  In reality,
because of the $\eta-\eta^\prime$ mixing,  $\eta_1$ ($\eta_8$) has additional component of $\eta$ ($\eta^\prime$).

Meson molecular states can be built from this PS nonet through the multiplication of $(\bm{1}_f\oplus\bm{8}_f)\otimes (\bm{1}_f\oplus\bm{8}_f$).
So, two-meson states can make several SU$_f$(3) multiplets like
\begin{eqnarray}
&&\underline{\text{PS}~\otimes \text{PS}} ~~~~ \text{\underline{two-meson multiplets}}\nonumber\\
&&(\bm{1}_f\otimes\bm{1}_f) \rightarrow ~\bm{1}^{\prime}\label{m1}\ , \\
&&(\bm{1}_f\otimes\bm{8}_f) \rightarrow ~\bm{8}^{\prime\prime}\label{m2}\ , \\
&&(\bm{8}_f\otimes\bm{8}_f) \rightarrow ~\bm{27} \oplus \bm{10}\oplus\bm{\overline{10}}\oplus \bm{8}\oplus \bm{8}^\prime \oplus \bm{1}\label{m3}\ .
\end{eqnarray}
Here we have suppressed the subscript ''$f$'' in denoting the two-meson multiplets in order to distinguish
them from the PS multiplets.
In this construction, there are two singlets ($\bm{1}^{\prime}, \bm{1}$),
and three octets ($\bm{8}^{\prime\prime},\bm{8}^\prime,\bm{8}$) that can be utilized to describe the two nonets in PDG.

From Eqs~(\ref{m1}),(\ref{m2}), we trivially obtain one molecular nonet ($\bm{1}^{\prime}$, $\bm{8}^{\prime\prime}$) whose meson compositions are given as,
\begin{eqnarray}
\bm{1}^{\prime}&=&\eta_1 \eta_1\label{ss}\ ,\\
(\bm{8}^{\prime\prime})^3_1 &=&\eta_1 K^+, (\bm{8}^{\prime\prime})^3_2 =\eta_1 K^0\label{o1}\ ,\\
(\bm{8}^{\prime\prime})^2_1 &=&\eta_1 \pi^+, \frac{1}{\sqrt{2}}[(\bm{8}^{\prime\prime})^1_1 - (\bm{8}^{\prime\prime})^2_2] =\eta_1 \pi^0 \label{o111}\ ,\\
(\bm{8}^{\prime\prime})^1_2 &=& \eta_1 \pi^-\label{o2}\ , \\
(\bm{8}^{\prime\prime})^3_3 &=&\eta_1 \eta_8\label{o3} ,\\
(\bm{8}^{\prime\prime})^2_3 &=&\eta_1 \bar{K}^0 ,(\bm{8}^{\prime\prime})^1_3 =\eta_1 K^-\label{o4}\ .
\end{eqnarray}
By construction, the singlet, $\bm{1}^{\prime}$, represents the molecular state of $\eta_1 \eta_1$
and the octet, $\bm{8}^{\prime\prime}$, represents two-meson states with the common constituent, $\eta_1$.
So all the modes in this molecular nonet contain the $\eta_1$ meson as a common constituent.
This nonet with this trivial structure is quite unlikely to represent either
of the two nonets.
(See Sec.~\ref{comparison} for further discussion.)

Another SU$_f$(3) nonet can be constructed from $\bf{8}, \bf{8}^\prime, \bm{1}$ in Eq.~(\ref{m3}).
But Eq.~(\ref{m3}) also has other multiplets like $\bm{27}$, $\bm{10}$, $\bm{\overline{10}}$.
So, even if this molecular nonet turns out to be physically feasible, additional explanations are still
needed as to why higher multiplets do not appear in the hadron spectrum.
Nevertheless, in this work, we investigate whether $\bf{8}, \bf{8}^\prime, \bm{1}$ in Eq.~(\ref{m3}) can
make a molecular nonet that can account for either of the two nonets in PDG.

To build a flavor nonet, we make use of the tensor method\footnote{For technical details in using the tensor notation, see Ref.~\cite{Oh:2004gz}.}
where all the SU$_f$(3) multiplets in Eq.~(\ref{m3}) are represented by appropriate tensors,
\begin{eqnarray}
\bm{27}^{ij}_{kl},\ \bm{10}_{ijk},\ \bm{\overline{10}}^{ij k},\ \bm{8}^i_j,\ (\bm{8}^\prime)^i_{j},\ \bm{1}\label{multiplets}\ .
\end{eqnarray}
Note, each tensor is symmetric under exchange of any two upper (or lower) indices [{\it e.g.},
$\bm{\overline{10}}^{ij k}=\bm{\overline{10}}^{jik}$,
$\bm{10}_{ijk}=\bm{10}_{jik}$], and traceless under the contraction of a upper and a lower
indices [ {\it e.g.}, $\bm{27}^{ij}_{il}=0$, $(\bm{8}^\prime)^i_i=0$ ].
At the same time, the $\bm{8}_f\otimes\bm{8}_f$ part in Eq.~(\ref{m3}) can be
written as
\begin{eqnarray}
P^i_{j} P^{i^\prime}_{j^\prime}\nonumber \ .
\end{eqnarray}

In the tensor method, the multiplets in Eq.~(\ref{multiplets}) appear in the group multiplication of Eq.~(\ref{m3}) because they
are possible tensors that can make SU$_f$(3) invariants with $P^i_{j} P^{i^\prime}_{j^\prime}$.
More concretely, tensors of the following forms,
\begin{eqnarray}
&&\bm{27}_{ii^\prime}^{jj^\prime},\ \epsilon_{ii^\prime k} \bm{\overline{10}}^{jj^\prime k},\ \epsilon^{jj^\prime k}\bm{10}_{ii^\prime k}\nonumber \\
&& \delta_i^{j^\prime} \bm{8}_{i^\prime}^j,\ \delta_{i^\prime}^j (\bm{8}^\prime)_i^{j^\prime},\ \delta_i^{j^\prime} \delta_{i^\prime}^j \bm{1}
            \label{tensor}\ ,
\end{eqnarray}
produce SU$_f$(3) invariants when multiplied by $P^i_j P^{i^\prime}_{j^\prime}$.
For instance, $\bm{27}^{jj^\prime}_{ii^\prime}P^i_j P^{i^\prime}_{j^\prime}$
forms an SU$_f$(3) invariant as all the indices are fully contracted so the $\bm{27}$ multiplet must be present in $P^i_{j} P^{i^\prime}_{j^\prime}$.
Another example, $\epsilon_{ii^\prime k} \bm{\overline{10}}^{jj^\prime k}P^i_j P^{i^\prime}_{j^\prime}$, also forms an SU$_f$(3) invariant
so the $\bm{10}$ multiplet must be present in $P^i_{j} P^{i^\prime}_{j^\prime}$ and so on.
Multiplets other than those in Eq.~(\ref{multiplets}), for example like $\bm{6}^{ij}, \bm{15}^{ij}_k$, do not appear in Eq.~(\ref{m3})
because they cannot make SU$_f$(3) invariants when multiplied with $P^i_{j} P^{i^\prime}_{j^\prime}$.
From each SU$_f$(3) invariant that Eq.~(\ref{tensor}) generates, one can then identify the two-meson states corresponding to each multiplet.

Since we want to make flavor nonets, we concentrate on two-meson multiplets of $\bm{1}$, $\bm{8}$, $\bm{8}^\prime$ in Eq.~(\ref{tensor}).
First, two-meson state for $\bm{1}$ is obtained by multiplying $\delta_i^{j^\prime} \delta^j_{i^\prime}$
on $P^i_j P^{i^\prime}_{j^\prime}$,
\begin{equation}
\bm{1}= \text{Tr}(P P)= \bm{\vec{\pi}}\cdot\bm{\vec{\pi}} + \overline{K}K + (\overline{K}K)^\dagger + \eta_8\eta_8 \label{singlet}\ .
\end{equation}
In the SU(3)$_f$ limit, this two-meson molecule, after normalized to the unity, corresponds to either $f_0(980)$ in the light nonet or
$f_0(1500)$ in the heavy nonet.

To find $\bm{8}$, we multiply $\delta^{j^\prime}_i \bm{8}^{j}_{i^\prime}$ on $P^i_j P^{i^\prime}_{j^\prime}$
to make an SU$_f$(3) invariant,
\begin{eqnarray}
\bm{8}^j_{i^\prime} P^i_j P^{i^\prime}_i = \bm{8}^j_{i^\prime} (P P)_j^{i^\prime}\ .
\end{eqnarray}
Since the octet ($\bm{8}$) that we are constructing is traceless, the term, $\frac{1}{3} \delta^{i^\prime}_j \text{Tr}(P P)$, can be inserted freely to
obtain
\begin{eqnarray}
\bm{8}^j_{i^\prime} (P P)_j^{i^\prime} = \bm{8}^j_{i^\prime} \left [(P P)_j^{i^\prime} - \frac{1}{3} \delta^{i^\prime}_j \text{Tr}(P P) \right ]\label{octet1}\ .
\end{eqnarray}
Eq.~(\ref{octet1}) can form a flavor singlet if the expression in the square bracket is identified as $\bm{8}_j^{i^\prime}$.
This observation leads us to the two-meson octet as
\begin{eqnarray}
\bm{8}^{i^\prime}_j = (P P)^{i^\prime}_j -\frac{1}{3} \delta^{i^\prime}_j \text{Tr}(P P)\label{octet}\ .
\end{eqnarray}

For the other octet, $\bm{8}^\prime$, we proceed similarly by multiplying $\delta^j_{i^\prime} (\bm{8}^{\prime})_i^{j^\prime}$ on $P^i_{j} P^{i^\prime}_{j^\prime}$, and
eventually find that $\bm{8}^\prime=\bm{8}$.  This result stems from the fact that we are multiplying the same pseudoscalar octet twice
to build meson-molecular states.
Thus, Eq.~(\ref{m3}) can make only one nonet composed of $\bm{1}$ [Eq.~(\ref{singlet})] and $\bm{8}$ [Eq.~(\ref{octet})].
Therefore, this molecular nonet cannot represent the two nonets in PDG.
Instead, this molecular nonet can represent only the light nonet or the heavy nonet.
Using the normalization for $P^i_j$ in Eq.~(\ref{normalization1}), we find that $\bm{8}^i_j$ is normalized according to
\begin{eqnarray}
\big ( \bm{8}^i_j, \bm{8}^k_l \big )=\frac{7}{3} \left ( \delta^{ik}\delta_{jl}-\frac{1}{3}\delta^i_j\delta^k_l\right ) \label{normalization2}\ .
\end{eqnarray}

Now, from Eq.~(\ref{octet}), it is straightforward to write down all the octet members explicitly in terms of two pseudoscalar mesons.
\begin{widetext}
\begin{eqnarray}
&&\bm{8}^3_1 =\frac{1}{\sqrt{2}}\pi^0 K^+ +\frac{1}{\sqrt{6}}\eta_8K^+ + \pi^+K^0 - \sqrt{\frac{2}{3}} K^+\eta_8\label{o31}\ ,\\
&&\bm{8}^3_2 =\pi^- K^+ - \frac{1}{\sqrt{2}}\pi^0 K^0 +\frac{1}{\sqrt{6}}\eta_8K^0 - \sqrt{\frac{2}{3}} K^0\eta_8\label{o32}\ ,\\
&&\bm{8}^2_1 =\frac{1}{\sqrt{2}}(\pi^0\pi^+ - \pi^+\pi^0)+\frac{1}{\sqrt{6}}(\eta_8\pi^+ + \pi^+\eta_8)+ K^+\bar{K}^0\label{o21}\ ,\\
\bm{8}^1_1 - &&\bm{8}^2_2 =\frac{1}{\sqrt{3}}\left [\pi^0\eta_8 + \eta_8 \pi^0 + \sqrt{3}(\pi^+\pi^-
- \pi^-\pi^+ + K^+ K^- - K^0\bar{K}^0) \right ]\label{o1122}\ ,\\
&&\bm{8}^1_2 = \frac{1}{\sqrt{2}}(\pi^-\pi^0 - \pi^0\pi^-)+\frac{1}{\sqrt{6}}(\eta_8\pi^- + \pi^-\eta_8)+ K^0K^-\label{o12}\ , \\
&&\bm{8}^3_3 =\frac{1}{\sqrt{3}}\left[ -\bm{\vec{\pi}}\cdot\bm{\vec{\pi}} - (\overline{K}K)^\dagger + 2 \overline{K}K + \eta_8\eta_8\right ]\label{o33}\ ,\\
&&\bm{8}^2_3 =K^-\pi^+ - \frac{1}{\sqrt{2}}\bar{K}^0\pi^0 +\frac{1}{\sqrt{6}}\bar{K}^0\eta_8
- \sqrt{\frac{2}{3}} \eta_8\bar{K}^0 \label{o23}\ ,\\
&&\bm{8}^1_3 =\frac{1}{\sqrt{2}}K^-\pi^0 + \bar{K}^0\pi^-   +  \frac{1}{\sqrt{6}} K^-\eta_8 - \sqrt{\frac{2}{3}} \eta_8 K^- \label{o13}\ .
\end{eqnarray}
\end{widetext}
Note that these states need to be normalized to the unity when they are matched to physical states that belong to the light nonet or the heavy nonet.

To give some justifications on these expressions, it is worth considering octet members, $\bm{8}^1_1, \bm{8}^2_2$,
whose expressions from Eq.~(\ref{octet}) are given as,
\begin{eqnarray}
\bm{8}^1_1&=&
\frac{1}{6}\pi^0\pi^0+\frac{2}{3}\pi^+\pi^--\frac{1}{3}\pi^-\pi^+ -\frac{1}{6}\eta_8\eta_8 \nonumber \\
&&+\frac{1}{\sqrt{12}}(\pi^0\eta_8+\eta_8\pi^0)  +\frac{2}{3} K^+K^- - \frac{1}{3} K^-K^+\nonumber \\
&& -\frac{1}{3} (K^0\bar{K}^0 + \bar{K}^0K^0) \label{o11}\ , \\
\bm{8}^2_2&=&\frac{1}{6}\pi^0\pi^0-\frac{1}{3}\pi^+\pi^- +\frac{2}{3}\pi^-\pi^+ -\frac{1}{6}\eta_8\eta_8 \nonumber \\
&&-\frac{1}{\sqrt{12}}(\pi^0\eta_8+\eta_8\pi^0)  -\frac{1}{3} (K^+K^- + K^-K^+)\nonumber \\
&& +\frac{2}{3} K^0\bar{K}^0 -\frac{1}{3} \bar{K}^0K^0 \label{o22}\ .
\end{eqnarray}
Because of the traceless condition, $\bm{8}^j_j=0$,
their sum, $\bm{8}^1_1+\bm{8}^2_2$, can be identified as an isoscalar member in the octet through
\begin{eqnarray}
\bm{8}^{3}_{3}=-[\bm{8}^1_1+\bm{8}^2_2]\label{iso1}\ ,
\end{eqnarray}
and its expression is neatly given as Eq.~(\ref{o33}).
One can also justify this identification of $\bm{8}^{3}_{3}$ by showing that
it is orthogonal to $\bm{1}$ in Eq.~(\ref{singlet}), $\langle \bm{1}| \bm{8}^{3}_{3}\rangle=0$.

In addition, the difference, $\bm{8}^1_1-\bm{8}^2_2$, which is clearly orthogonal to $\bm{8}^{3}_{3}$ in Eq.~(\ref{iso1}),
can be identified as an isovector member whose expression is given by Eq.~(\ref{o1122}).
This identification can be further justified by showing that the isospin ladder operators, $I_{\pm}$, when applied to Eqs.~(\ref{o21}),(\ref{o12}),
give
\begin{eqnarray}
&&I_{-} \bm{8}^2_1\propto -[\bm{8}^1_1 - \bm{8}^2_2]\ ,\\
&&I_{+} \bm{8}_2^1\propto [\bm{8}^1_1 - \bm{8}^2_2]\ .
\end{eqnarray}
So, it is clear that Eqs.~(\ref{o21}), (\ref{o1122}) and (\ref{o12}) form an isospin triplet.
\begin{figure}[t]
\centering
\bigskip
\epsfig{file=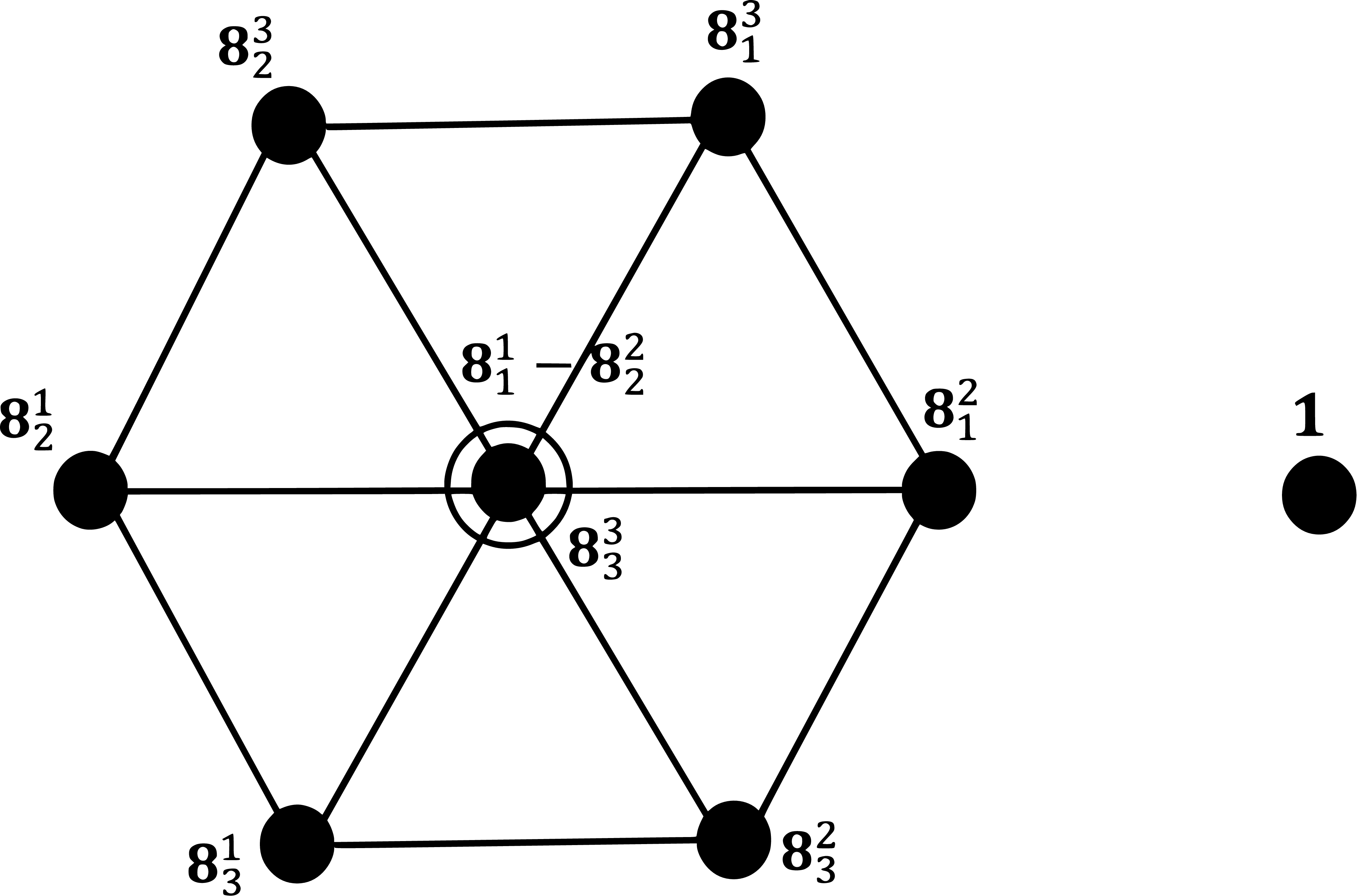, width=0.6\hsize}
\caption{Weight diagram for the nonet, $\bm{1}\oplus \bm{8}$.}
\label{weight}
\end{figure}
Weight diagram for this nonet is given in Fig.~\ref{weight} so one can assign $\bm{1}$ and $\bm{8}^i_j$ easily
to the members of the light nonet or to the members of the heavy nonet.
To make our presentation clear, this assignment has been listed in Table~\ref{correspondence}.

A few comments are in order. Two-meson modes in Eqs.~(\ref{o31})$\cdot\cdot\cdot$~(\ref{o13})
represent possible decay modes that can be measured if the decays are kinematically allowed.
These modes are clearly different from two-meson modes in Eqs.~(\ref{doublet700})~$\cdot\cdot\cdot$~(\ref{1500mode})
obtained from the tetraquark mixing model.
Thus, as advertised, two-meson modes can be used to distinguish the tetraquark mixing model and meson molecules.
Another thing to point out is that a term like $\pi^0\pi^+$ in Eq.~(\ref{o21}) should be treated independently from $\pi^+\pi^0$
so their combination like $\pi^0\pi^+ - \pi^+\pi^0$ should not vanish. In fact, this pion combination makes an isospin state of $I=1, I_z=1$
that corresponds to the isospin of the two-meson state, $\bm{8}^2_1$.
Similar terms in the other isovector members, Eqs.~(\ref{o1122}),(\ref{o12}), should not vanish also.
To put it other way,
if such terms vanish, the isovector members in $\bm{8}^i_j$ of Eq.~(\ref{octet}) are no longer normalized according to Eq.~(\ref{normalization2}).

\begin{table}[t]
\centering
\begin{tabular}{c|crr|c|c}\hline
molecular nonet& $I$ & $I_z$& $Y$& light nonet &  heavy nonet  \\
\hline 
$\bm{1}$ & $0$ & $0$ & $0$ & $f_0(980)$  & $f_0(1500)$   \\[1mm]
$\bm{8}^3_1$ & $\frac{1}{2}$ & $\frac{1}{2}$ & $1$ & $K_0^{*+}(700)$  & $K_0^{*+}(1430)$   \\[1mm]
$\bm{8}^3_2$ & $\frac{1}{2}$ & $-\frac{1}{2}$ & $1$ & $K_0^{*0}(700)$  & $K_0^{*0}(1430)$   \\[1mm]
$\bm{8}^2_1$ & $1$ & $1$ & $0$ & $a^+_0(980)$  & $a^+_0(1450)$   \\[1mm]
$\frac{1}{\sqrt{2}}[\bm{8}^1_1 - \bm{8}^2_2]$ & $1$ & $0$ & $0$ &  $a^0_0(980)$  & $a^0_0(1450)$   \\[1mm]
$\bm{8}^1_2$ & $1$ & $-1$ & $0$ &  $a^-_0(980)$  & $a^-_0(1450)$   \\[1mm]
$\bm{8}^3_3$ & $0$ & $0$ & $0$ & $f_0(500)$  & $f_0(1370)$   \\
$\bm{8}^2_3$ & $\frac{1}{2}$ & $\frac{1}{2}$ & $-1$ & $\bar{K}_0^{*0}(700)$  & $\bar{K}_0^{*0}(1430)$   \\[1mm]
$\bm{8}^1_3$ & $\frac{1}{2}$ & $-\frac{1}{2}$ & $-1$ &  $K_0^{*-}(700)$  & $K_0^{*-}(1430)$   \\[1mm]
\hline
\end{tabular}
\caption{Molecular nonet in Eqs.~(\ref{o31})$\cdot\cdot\cdot$~(\ref{o13}) that can assigned to the light nonet or the heavy nonet
after normalized to the unity.
Also given are isospin, its z-component, and hypercharge of each member.}
\label{correspondence}
\end{table}

\section{Comparison between tetraquark mixing model and meson molecules}
\label{comparison}

Up to now, we have presented two approaches, the tetraquark mixing model and meson molecules,
to describe the two nonets in PDG:
the light nonet [$a_0 (980)$, $K_0^* (700)$, $f_0 (500)$, $f_0 (980)$] and
the heavy nonet [$a_0 (1450)$, $K_0^* (1430)$, $f_0 (1370)$, $f_0 (1500)$].
In this section, we compare the two approaches and examine
which approach is more appropriate to describe the two nonets.

In the tetraquark mixing model, the two nonets in PDG are described by the mixing formulas, Eqs.~(\ref{heavy}),(\ref{light}).
In this approach, it is important to introduce the two tetraquark nonets, $|000\rangle,|011\rangle$, because the successful aspects
of this model come mostly from the fact that the two types of tetraquarks mix with each other. This also means that
the two physical nonets should not be treated separately.

In meson molecules, it is also possible to build two molecular nonets from two PS mesons.
One molecular nonet is ($\bm{1}^\prime$, $\bm{8}^{\prime\prime}$) constructed from Eqs.~(\ref{m1}),(\ref{m2})
and the other nonet is ($\bm{1}$, $\bm{8}$) from Eq.~(\ref{m3}).
The first nonet, ($\bm{1}^\prime$, $\bm{8}^{\prime\prime}$), cannot represent either of the two nonets in PDG by two reasons.
One reason is that this nonet cannot reproduce the inverted mass ordering satisfied by the two nonets~\cite{Kim:2016dfq}.
This nonet, whose meson compositions are given in Eqs.~(\ref{ss})~$\cdot\cdot\cdot$~(\ref{o4}), has
$\eta_1$ as a common constituent. This nonet has the mass hierarchy driven solely by the PS mesons, $M(\pi)<M(K)<M(\eta_8)$,
which is opposite to the inverted mass ordering.
Another reason is that the meson compositions, in Eqs.~(\ref{ss})~$\cdot\cdot\cdot$~(\ref{o4}),
are not consistent with the decay modes of the two physical nonets.
For example, $(\bm{8}^{\prime\prime})^3_1$ in Eq.~(\ref{o1}), which can be matched to $K^*_0(700)$
in the light nonet or $K^*_0(1430)$ in the heavy nonet, has the meson
composition of $K^+\eta_1$ only. This composition is not consistent
with the experimental fact that $K^*_0(700)$ or $K^*_0(1430)$ decays mostly to $K\pi$.
Another molecular nonet is composed of $\bm{1}$ [Eq.~(\ref{singlet})] and $\bm{8}^i_j$ [Eqs.~(\ref{o31})~$\cdot\cdot\cdot$~(\ref{o13})].
This nonet has non-trivial meson compositions so it can be tested
either for the light nonet or for the heavy nonet. But this molecular nonet cannot describe both nonets simultaneously.
Therefore, there are limitations in describing the physical two nonets with the meson molecular model.

Still, one may pursue a specific mixing scheme that combine the two molecular nonets.
Note that the two-meson modes from the trivial nonet, ($\bm{1}^\prime$, $\bm{8}^{\prime\prime}$),
have the $\eta_1$ state as a common constituent.
Through the $\eta-\eta^\prime$ mixing, $\eta_1$ as well as $\eta_8$ has the meson components, $\eta$ and $\eta^\prime$.
If a certain mechanism is invoked to mix ($\bm{1}^\prime$, $\bm{8}^{\prime\prime}$)
in Eqs.~(\ref{ss})~$\cdot\cdot\cdot$~(\ref{o4}) and ($\bm{1}$, $\bm{8}$) in Eq.~(\ref{singlet}), Eqs.~(\ref{o31})~$\cdot\cdot\cdot$~(\ref{o13}),
this mixing should occur only through the two-meson modes containing $\eta_1$ or $\eta_8$.
This mixing, therefore, cannot affect other two-meson modes
like $\overline{K}K$, $\pi K$, $\bm{\vec{\pi}}\cdot\bm{\vec{\pi}}$ etc.
Because of this, this mixing cannot be universal and affects only the two-meson modes involving $\eta_1$, $\eta_8$.
Therefore, it is unlikely that this molecular model reproduces phenomenological
aspects that apply universally to all the members in each nonet.
For example, the huge mass gap, around $\Delta M\approx 500$ MeV, that universally exists
between the two nonets may not be reproduced by a mixing scheme that affects only the two-meson modes involving $\eta_1$, $\eta_8$.
Moreover, it certainly cannot reproduce the striking prediction of the tetraquark mixing model, namely, the coupling strengths that
are universally enhanced in the light nonet but suppressed in the heavy nonet as shown in Eqs.~(\ref{doublet700})~$\cdot\cdot\cdot$~(\ref{1500mode}).
Since this prediction is supported by the experimental partial widths with the trend of Eq.~(\ref{trend}),
this is one of strong indications that the tetraquark mixing model is more appropriate to describe the two nonets in PDG.

To test ($\bm{1}$, $\bm{8}$) further,
we compare the two-meson modes of $(\bm{1}, \bm{8})$ in Eq.~(\ref{singlet}), Eqs.~(\ref{o31})~$\cdot\cdot\cdot$~(\ref{o13})
with those in Eqs.~(\ref{doublet700})~$\cdot\cdot\cdot$~(\ref{1500mode}) from the tetraquark mixing model.
Since ($\bm{1}$, $\bm{8}$) can represent only one of the two nonets,
we will first treat ($\bm{1}$, $\bm{8}$) as the heavy nonet and later we discuss the case when this molecular nonet is interpreted
as the light nonet.  Here, for instance,
we take the isodoublet member, $K_0^{*+}(1430)$, and compare the two-meson modes in Eq.~(\ref{o31}) for $\bm{8}^3_1$
with the corresponding modes in Eq.~(\ref{doublet1430}) from the tetraquark mixing model.
But, our analysis can be applied to all members similarly.

There are two things that can differentiate the two equations. Eq.~(\ref{doublet1430}) and Eq.~(\ref{o31}) have the same two-meson components
like $\pi^0 K^+$, $\pi^+ K^0$, $\eta_8 K^+$, $K^+\eta_8$ but their relative coefficients are different in some modes.
In Eq.~(\ref{doublet1430}), the two modes, $K^+\eta_8$ and $\eta_8 K^+$, have the same coefficient but
in Eq.~(\ref{o31}), the $K^+\eta_8$ mode has the coefficient twice of the $\eta_8 K^+$ mode.
This difference, however, cannot be measured because the $K^+\eta_8$ and $\eta_8 K^+$ are not counted as separate modes in experiments.

Another difference that can distinguish the two approaches
is that the $K^+\eta_1$ mode, which appears in Eq.~(\ref{doublet1430}), is missing in Eq.~(\ref{o31}).
The $K^+\eta_1$ mode is missing in $\bm{8}^3_1$ [Eq.~(\ref{o31})] because its octet, $\bm{8}^i_j$, was constructed from Eq.~(\ref{m3})
where the $\eta_1$ state does not participate to begin with.
If the missing mode, $K^+\eta_1$, is interpreted as the absence of the $K^+\eta^\prime$ mode, this can be used to
advocate the tetraquark mixing model because the appearance of the $K^+\eta^\prime$ mode is supported by the experimental
decay modes of $K_0^{*}(1430)$~\cite{PDG22}.
However, the $K^+\eta^\prime$ mode can appear also from the $K^+\eta_8$ mode in Eq.~(\ref{o31})
as it can make small $K^+\eta^\prime$ mode from the $\eta-\eta^\prime$ mixing.
Experimentally, three modes, $\pi K$, $\eta K$, $K\eta^\prime$, have been reported as the decay modes of $K_0^{*}(1430)$,
and it is acknowledged that both approaches, the tetraquark mixing model [Eq.~(\ref{doublet1430})] and the meson molecular model [Eq.~(\ref{o31})],
can predict these experimental modes but with different branching ratios.
Measuring the branching ratios could be one possible way to differentiate between the two models, although its feasibility
is currently questionable due to the lack of experimental branching ratio for the $K\eta^\prime$ mode in the $K_0^{*}(1430)$ decay modes.

One can try to interpret $\bm{8}^3_1$ in Eq.~(\ref{o31}) as a member of the light nonet, $K_0^{*}(700)$,
and compare this molecular state with the two meson modes in Eq.~(\ref{doublet700}), which has the same modes as in Eq.~(\ref{doublet1430}).
In this case, the distinction between the two approaches becomes more obscure.
Experimentally, $K_0^{*}(700)$ has one decay mode, $\pi K$, and both approaches have this mode as one can
see in Eq.~(\ref{doublet700}) and Eq.~(\ref{o31}).
Other two-meson modes, $K\eta_8$ and $K\eta_1$, which can distinguish between Eq.~(\ref{doublet700}) and Eq.~(\ref{o31}),
cannot be measured experimentally due to the kinematical constraint.

Similar situations occur for other members when one compares the
meson molecular model of Eqs.~(\ref{o31})~$\cdot\cdot\cdot$~(\ref{o13}) with two-meson modes of Eqs.~(\ref{doublet700})~$\cdot\cdot\cdot$~(\ref{1500mode})
in the tetraquark mixing model.
A separate comparison can be made depending on whether the meson molecules are interpreted as the light nonet or the heavy nonet.
In this consideration, there are also various examples that have different branching ratios depending on the two approaches.
For example, in Eq.~(\ref{o21}), the $\pi^+\eta_8$ mode has the coefficient, $1/\sqrt{6}$, relative to the $K^+\bar{K}^0$ mode
but, in Eq.~(\ref{isovector980}), the $\pi^+\eta_8$ mode has the different coefficient of $\sqrt{2/3}$ relative to the $K^+\bar{K}^0$ mode.
However, all these investigations based on the relative branching ratios
need to be deferred to the future until the theoretical and experimental situations become more reliable.

Instead, from this comparison study, we report one clear distinction that can be seen from the isovector members.
For the isovector resonance, either $a_0^{+}(980)$ or $a_0^{+}(1450)$, we
compare its two-meson modes in Eq.~(\ref{o21}) [$\bm{8}^2_1$] from the molecular model with those from the tetraquark mixing model
as in Eq.~(\ref{isovector980}) or Eq.~(\ref{isovector1450}).
The two-pion mode, $\pi^0\pi^+ - \pi^+\pi^0$, appears in Eq.~(\ref{o21}) but missing in Eq.~(\ref{isovector980}) or Eq.~(\ref{isovector1450}).
The similar distinction can be expected from other isovector members, $a_0^0, a_0^-$.
As we mentioned already, in the meson molecular model,
this two-pion mode, $\pi^0\pi^+ - \pi^+\pi^0$, makes an isospin state of $I=1,I_z=1$ that corresponds to
the isospin state of $a_0^{+}(1450)$ or $a_0^{+}(980)$. The two-pion mode is also necessary in order
to maintain the normalization of Eq.~(\ref{normalization2}) consistently with the other octet members of $\bm{8}^i_j$.
So its presence in the meson molecular model seems natural.
At the same time, in the tetraquark mixing model, it is also natural that this mode, $\pi^0\pi^+ - \pi^+\pi^0$, does not appear in the two-meson modes of
$a_0^{+}(980)$ in Eq.~(\ref{isovector980})  [or $a_0^{+}(1450)$ in Eq.~(\ref{isovector1450})].
Both resonances, $a_0^{+}(1450)$ and $a_0^{+}(980)$, have the same flavor structure of $(su-us)(\bar{d}\bar{s}-\bar{s}\bar{d})$
in the tetraquark mixing model and, therefore, they cannot fall-apart into two-pion states that have no strange quarks ($s,\bar{s}$) in the final states.
In this regard, the two-pion mode can clearly distinguish the two approaches.
Experimentally, $a_0 (980)$, $a_0 (1450)$, do not have the two-pion modes even though these modes are
energetically possible from $a_0 (980)$, $a_0 (1450)$.  Therefore, the tetraquark mixing model is supported by the experimental data.
This can be another indication that the tetraquark mixing model is more appropriate to describe the two nonets in PDG.

\section{Summary}
\label{sec:summary}

In this work, we have examined the tetraquark mixing model and meson molecules in describing the two nonets in the $J^P=0^+$ channel.
The tetraquark mixing model that have been proposed and tested in various occasions in Refs.~\cite{Kim:2016dfq, Kim:2017yvd, Kim:2018zob,Kim:2018zob,Lee:2019bwi,Kim:2017yur,Kim:2022qfj,Kim:2019mbc}
has some successful features such as reproducing qualitatively
the masses of the two nonets and the mass difference between them.
Most notably, the mixing model predicts that the coupling strengths
into two PS mesons are enhanced in the light nonet and suppressed in the heavy nonet.
This prediction is indeed supported by the experimental partial decay widths.
To show this more explicitly, we have presented two-meson modes from the tetraquark mixing model.
All these successful aspects stem from the fact that the two tetraquark types that form two flavor nonets
mix with each other when creating the two physical nonets in PDG.

As an alternative description other than tetraquarks, we have constructed SU$_f$(3) molecular nonets by combining two PS mesons.
It is also possible to make two flavor nonets from this meson molecular model.
But one of them forms a trivial nonet whose meson compositions are not consistent with the mass ordering and the decay modes of the two nonets in PDG.
The second molecular nonet has a non-trivial structure but it can be tested for one nonet only, either the light nonet or the heavy nonet.
Therefore, it is difficult to describe the two nonets altogether by the meson molecular model.
Accordingly, this molecular model cannot reproduce
successful aspects of the tetraquark mixing model such as the mass splitting between
the two nonets, and the enhancement or suppression of the coupling strengths.
To test further whether the second molecular nonet is physically feasible,
we compare its two-meson modes
with those from the tetraquark mixing model.
Some of two-meson modes are found to have different branching ratios depending on the two approaches.
In principle, we could use the branching ratios to determine which model is more realistic,
but we defer actual calculations until the theoretical and experimental situations become more reliable.

However, there is one clear distinction that can distinguish the two approaches in the isovector resonances.
In the isovector channel, the two-pion modes appear in the meson molecular model but they are absent
in the tetraquark mixing model.
The absence of the two-pion modes is supported by the experimental decay modes of the isovector resonances.
We believe that this is another indication that the tetraquark mixing model is more appropriate to describe
the two nonets in PDG.

\acknowledgments

This work was supported by the National Research Foundation of Korea(NRF) grant funded by the
Korea government(MSIT) (No. NRF-2023R1A2C1002541, No. NRF-2018R1A5A1025563).


\begin{thebibliography}{10}


\bibitem{Belle03}
S.~K.~Choi \textit{et al.} [Belle],
Phys. Rev. Lett. \textbf{91}, 262001 (2003).

\bibitem{LHCb:2016axx}
R.~Aaij \textit{et al.} [LHCb],
Phys. Rev. Lett. \textbf{118}, no.2, 022003 (2017).

\bibitem{BESIII:2013ouc}
M.~Ablikim \textit{et al.} [BESIII],
Phys. Rev. Lett. \textbf{111}, no.24, 242001 (2013).



\bibitem{Xiao:2013iha}
T.~Xiao, S.~Dobbs, A.~Tomaradze and K.~K.~Seth,
Phys. Lett. B \textbf{727}, 366-370 (2013).

\bibitem{LHCb:2021auc}
R.~Aaij \textit{et al.} [LHCb],
Nature Commun. \textbf{13}, no.1, 3351 (2022).

\bibitem{LHCb:2021vvq}
R.~Aaij \textit{et al.} [LHCb],
Nature Phys. \textbf{18}, no.7, 751-754 (2022).


\bibitem{LHCb:2019kea}
R.~Aaij \textit{et al.} [LHCb],
Phys. Rev. Lett. \textbf{122}, no.22, 222001 (2019).

\bibitem{LHCb:2015yax}
R.~Aaij \textit{et al.} [LHCb],
Phys. Rev. Lett. \textbf{115}, 072001 (2015).



\bibitem{Jaffe77a}
  R.~L.~Jaffe,
  Multiquark hadrons. 1. The Phenomenology of $Q\bar{Q}^2$ mesons,
  Phys. Rev. D {\bf 15}, 267 (1977).


\bibitem{Jaffe77b}
  R.~L. Jaffe,
  Multiquark hadrons. 2. Methods,
  Phys.\ Rev.\ D {\bf 15}, 281 (1977).


\bibitem{Jaffe04}
  R.~L.~Jaffe,
  Exotica,
  Phys. Rept.  {\bf 409}, 1 (2005).


\bibitem{Kim:2016dfq}
H.~Kim, M.~K.~Cheoun and K.~S.~Kim,
Eur. Phys. J. C \textbf{77}, no.3, 173 (2017)
[erratum: Eur. Phys. J. C \textbf{77}, no.8, 545 (2017)].

\bibitem{Kim:2017yvd}
H.~Kim, K.~S.~Kim, M.~K.~Cheoun and M.~Oka,
Phys. Rev. D \textbf{97}, no.9, 094005 (2018).


\bibitem{Kim:2018zob}
H.~Kim, K.~S.~Kim, M.~K.~Cheoun, D.~Jido and M.~Oka,
Phys. Rev. D \textbf{99}, no.1, 014005 (2019).


\bibitem{Lee:2019bwi}
H.~J.~Lee, K.~S.~Kim and H.~Kim,
Phys. Rev. D \textbf{100}, no.3, 034021 (2019).


\bibitem{Kim:2017yur}
K.~S.~Kim and H.~Kim,
Eur. Phys. J. C \textbf{77}, no.7, 435 (2017).


\bibitem{Kim:2022qfj}
H.~Kim and K.~S.~Kim,
Eur. Phys. J. C \textbf{82}, no.12, 1113 (2022).


\bibitem{Kim:2019mbc}
H.~Kim,
XVIII Int. Conf. on Hadron Spectroscopy (HADRON2019),
doi:10.1142/9789811219313\_0025
[arXiv:1911.09904 [hep-ph]].

\bibitem{Guo:2017jvc}
F.~K.~Guo, C.~Hanhart, U.~G.~Mei\ss{}ner, Q.~Wang, Q.~Zhao and B.~S.~Zou,
Rev. Mod. Phys. \textbf{90}, no.1, 015004 (2018)
[erratum: Rev. Mod. Phys. \textbf{94}, no.2, 029901 (2022)].

\bibitem{Maiani:2004vq}
L.~Maiani, F.~Piccinini, A.~D.~Polosa and V.~Riquer,
Phys.\ Rev.\ D \textbf{71}, 014028 (2005).


\bibitem{Kim:2016tys}
H.~Kim, K.~S.~Kim, M.~K.~Cheoun, D.~Jido and M.~Oka,
Eur. Phys. J. A \textbf{52}, no.7, 184 (2016).

\bibitem{Tornqvist:2004qy}
N.~A.~Tornqvist,
Phys. Lett. B \textbf{590}, 209-215 (2004).

\bibitem{Tornqvist:1993ng}
N.~A.~Tornqvist,
Z. Phys. C \textbf{61}, 525-537 (1994).

\bibitem{Du:2019pij}
M.~L.~Du, V.~Baru, F.~K.~Guo, C.~Hanhart, U.~G.~Mei\ss{}ner, J.~A.~Oller and Q.~Wang,
Phys. Rev. Lett. \textbf{124}, no.7, 072001 (2020).

\bibitem{Xiao:2019mvs}
C.~J.~Xiao, Y.~Huang, Y.~B.~Dong, L.~S.~Geng and D.~Y.~Chen,
Phys. Rev. D \textbf{100}, no.1, 014022 (2019).


\bibitem{WASA-at-COSY:2011bjg}
P.~Adlarson \textit{et al.} [WASA-at-COSY],
Phys. Rev. Lett. \textbf{106}, 242302 (2011).

\bibitem{Kim:2020rwn}
H.~Kim, K.~S.~Kim and M.~Oka,
Phys. Rev. D \textbf{102}, no.7, 074023 (2020).


\bibitem{Dyson:1964xwa}
F.~Dyson and N.~H.~Xuong,
Phys. Rev. Lett. \textbf{13}, no.26, 815-817 (1964).


\bibitem{Janssen:1994wn}
G.~Janssen, B.~C.~Pearce, K.~Holinde and J.~Speth,
Phys. Rev. D \textbf{52}, 2690-2700 (1995).


\bibitem{Weinstein:1990gu}
J.~D.~Weinstein and N.~Isgur,
Phys. Rev. D \textbf{41}, 2236 (1990).


\bibitem{Branz:2007xp}
T.~Branz, T.~Gutsche and V.~E.~Lyubovitskij,
Eur. Phys. J. A \textbf{37}, 303-317 (2008).

\bibitem{Branz:2008ha}
T.~Branz, T.~Gutsche and V.~E.~Lyubovitskij,
Phys. Rev. D \textbf{78}, 114004 (2008).

\bibitem{Ahmed:2020kmp}
H.~A.~Ahmed and C.~W.~Xiao,
Phys. Rev. D \textbf{101}, no.9, 094034 (2020).


\bibitem{Molina:2008jw}
R.~Molina, D.~Nicmorus and E.~Oset,
Phys. Rev. D \textbf{78}, 114018 (2008).


\bibitem{PDG22}
R.~L.~Workman [Particle Data Group],
Review of Particle Physics,
PTEP {\bf 2022}, 083C01 (2022).


\bibitem{Oh:2004gz}
  Yongseok~Oh and Hungchong~Kim,
  Phys.\ Rev.\ D {\bf 70}, 094022 (2004).

\end{thebibliography}
\end{document}